\documentclass[twocolumn, 10pt]{article}

\usepackage[utf8]{inputenc}
\usepackage[english]{babel}
\usepackage[margin=1.8cm]{geometry}
\usepackage{authblk} 
\usepackage{amsmath, amssymb, amsfonts}
\usepackage{graphicx}
\usepackage{hyperref}
\usepackage{cite}
\usepackage{booktabs}
\usepackage{abstract}
\usepackage{float}
\usepackage{tikz}
\usetikzlibrary{arrows.meta, positioning, calc}

\hypersetup{
    colorlinks=true,
    linkcolor=blue,
    urlcolor=cyan,
    citecolor=blue,
}

\title{\textbf{Three-Dimensional Volumetric Reconstruction of Native Chilean Pollen via Lens-Free Digital In-line Holographic Microscopy}}

\author[1]{J. Staforelli-Vivanco \thanks{Corresponding author: jstaforelli@udec.cl}}
\author[1]{V. Salamanca-Levi}
\author[1]{B. Muñoz-Cepeda}
\author[1]{R. Jofré-Cerda}
\author[1]{C. Valdés-Cruces}
\author[2]{M. Rondanelli-Reyes}
\author[2]{I. Lamas}
\author[3]{J. M. Troncoso}

\affil[1]{Departamento de Física, Facultad de Ciencias Físicas y Matemáticas, Universidad de Concepción, Concepción, Chile.}
\affil[2]{Laboratory of Palynology and Plant Ecology, School of Sciences and Technologies, University of Concepción, Los Angeles Campus}
\affil[3]{Universidad Arturo Prat, Sede Victoria, Chile}

\date{\today}

\begin{document}

\maketitle

\begin{abstract}
This study presents a methodology for the three-dimensional (3D) volumetric reconstruction and morphological characterization of native Chilean pollen grains using a lens-free Digital In-line Holographic Microscopy (DLHM) system. A 532 nm point-source configuration and a 3.45 $\mu$m pixel pitch CMOS sensor, with a source--sample distance of $0.46$ mm and a sample--sensor stand-off of 25 mm, give a geometric magnification of $\simeq 55\times$ and an effective pixel of 62.9 nm at the object plane. The complex wavefronts of \textit{Anthemis cotula}, \textit{Gevuina avellana} and \textit{Conium maculatum} were reconstructed via the Kirchhoff-Helmholtz transform, yielding volumes from $3780.2 \pm 18$ to $4320.5 \pm 15$ $\mu$m$^3$ and identifying \textit{A. cotula} as the least spherical species ($\Psi = 0.76 \pm 0.03$) owing to its echinate exine, against $0.89 \pm 0.02$ for \textit{G. avellana}. Three independent targets convert these reconstructions into traceable measurements. Certified polystyrene microspheres ($\varnothing = 20.0$ $\mu$m) make the source--sample distance an optically retrieved quantity: the 2.5D height map reproduces a hemispherical cap of $2068$ $\mu$m$^3$ against an analytical $2094.4$ $\mu$m$^3$, a deviation of $1.3\%$. A human erythrocyte, reconstructed with a haemoglobin refractive-index increment taken from the literature rather than fitted, returns $88.9$ fL with a rim height of $2.36$ $\mu$m and a central dimple of $1.03$ $\mu$m, all within physiological range. Finally, the Transport-of-Intensity Equation applied to a through-focus triplet from a separate bright-field microscope (Plan Fluor $60\times$ with a $0.55\times$ reduction lens, $0.073$ $\mu$m object-plane sampling calibrated against a USAF 1951 target) reproduces the projected morphology and phase topography of \textit{G. avellana} on an instrument sharing neither optics, illumination nor inversion algorithm with the holographic channel, and returns an area consistent in order of magnitude with the value reported for thousands of expert-verified masks of the same species. No volume is derived from that channel, since its defocus is an instrumental step unrelated to the optical thickness of the grain. These results support label-free "digital fingerprints" for automated melissopalynology in a South American biodiversity hotspot.
\end{abstract}

\section{Introduction}

The accurate identification of pollen grains is a fundamental task in melissopalynology for several reasons, particularly botanic, ecologic, and economic such as honey certification and fraud prevention. In Chile, the authentication of endemic
honeys is vital for international trade. Traditional microscopy is labor-intensive; thus, polen digitization is essential for future automatization and deep learning models. However, existing datasets like Pollen13K primarily feature European species and very few repositories in 3D visualization exist \cite{Batiato2020}. 

Since the primary concepts in Holography \cite{Goodman2005} and digital holographic microscopy \cite{Vanlingten1966, Lee2007,Grier2007,Kim2011} Digital Lens-less Holographic Microscopy (DLHM) has emerged as a transformative technique for biological imaging, bypassing the diffraction limits and aberrations of traditional refractive optics. By capturing the phase information encoded in the interference between a reference wave and the light scattered by a specimen, DLHM enables a label-free 3D reconstruction of micro-scale objects \cite{Martin2022,Lopera2024,Acosta2024,Buitrago2023}. 

Up to our knowledge, very few studies have reported the use of holographic microscopy in pollen grains, particularly with the lensless technique and with pollen targets from South America. Recent research by Kumar et al. (2023) has demonstrated that quantitative phase imaging via DHM can accurately assess pollen viability without traditional staining procedures \cite{Kumar2023}. In aerosols, Sauvageat et al. describe a method to measure density of polen using an in-line Digital Holography system and Van Hout et al. present the validation of an operational automatic pollen monitoring system based on digital holography \cite{Sauvageat2020,VanHout2004}. Warshaneyan et al. (2025), explores the application of deep learning to improve and automate pollen grain detection and classification in both optical and holographic microscopy images, with focus on veterinary cytology use cases \cite{Warshaneyan2025}. This aligns with advancements in physics-driven neural networks for single-shot morphology retrieval \cite{kim2025,oconnor2020}.

In the context of palynology, the ability to resolve the intricate exine structures of pollen grains without chemical staining is paramount for high-throughput environmental analysis. This work settles the proof of principle to fill this gap.

A recurrent limitation of lens-free geometries is metrological rather than optical. Since magnification is purely geometric, $I_M = L/z$, every reported length, area and volume is only as accurate as the estimate of the source--sample distance $z$, hard to gauge mechanically in a compact assembly. We replace that estimate by an optical one, using certified polystyrene microspheres as an internal standard: a sphere of known diameter and index is at once a lateral ruler, an axial ruler and a volumetric ground truth, so a single frame fixes $z$, the effective pixel $\Delta x_{obj}$ and the height scale \cite{GarciaSucerquia2006,Sultanova2009}. The same logic is extended to a biological target of well-documented morphology, the human erythrocyte, whose thickness profile and corpuscular volume are tabulated and whose refractive index follows from the intracellular haemoglobin concentration \cite{Gienger2019,Evans1972}.

Any validation performed inside the holographic pipeline, however, tests only internal consistency, since an error in the axial scale would propagate identically through calibration and measurement. To break this circularity we add a second, independent channel based on the Transport-of-Intensity Equation (TIE) \cite{Teague1983,Streibl1984,Paganin1998,Zuo2020,Wang2026}. Crucially, the intensity planes fed to the solver are \emph{not} numerical reconstructions of a hologram: they are through-focus bright-field micrographs whose optics, illumination, detector and sampling all differ from the DLHM channel, on a lateral scale traceable to a resolution standard.

\section{Mathematical Formalism}

The core of our reconstruction process lies in the numerical propagation of the complex wavefront. Given a point-source configuration, the reconstruction of the object wave $U(\mathbf{r})$ at a distance $z$ from the source is governed by the Kirchhoff-Helmholtz integral.

For a digital hologram $I(\xi, \eta)$ recorded at the detector plane (at a distance $L$ from the source), the reconstructed wavefront $K(\mathbf{r})$ is calculated as:

\begin{equation}
K(\mathbf{r}) = \iint_{\Gamma} I(\mathbf{\xi}) \exp \left( \frac{ik \mathbf{r} \cdot \mathbf{\xi}}{|\mathbf{\xi}|} \right) d^2\xi
\end{equation}

In practice, we implement a discrete version of this transform adapted for spherical wave illumination. The implementation utilizes the algorithms developed by Latychevskaia and Fink \cite{latychevskaia2015}, which allow for the retrieval of both intensity and phase by treating the hologram as a 2D projection of the 3D scattering potential. The complex amplitude at any plane $z$ is given by:

\begin{equation}
U(x, y, z) = \frac{1}{i\lambda} \iint I(\xi, \eta) \frac{\exp(ik\rho)}{\rho} \cos \theta \, d\xi d\eta
\label{eq:kh_propagation}
\end{equation}

where $\rho = \sqrt{(x-\xi)^2 + (y-\eta)^2 + z^2}$ represents the distance between the hologram and the reconstruction plane, and $\cos \theta$ is the obliquity factor.

\subsection{Transport-of-intensity phase retrieval}
\label{sec:tie_theory}

The KH transform recovers the complex field from a single hologram, but in an in-line geometry the reconstruction is contaminated by the conjugate (twin) image, which biases the phase near strongly scattering structures such as the echinate exine; a validation carried out on the reconstructed stack would therefore inherit the very errors it is meant to detect. To obtain an external reference we solve the Transport-of-Intensity Equation (TIE), formulated by Teague as a deterministic alternative to interferometric phase measurement \cite{Teague1983,Streibl1984}, on intensity images recorded with a separate bright-field microscope. For a paraxial monochromatic field the conservation of energy along the optical axis links the axial derivative of the irradiance to the transverse phase gradient:

\begin{equation}
-k\,\frac{\partial I(x,y,z)}{\partial z} = \nabla_{\perp} \cdot \left[ I(x,y,z)\, \nabla_{\perp} \varphi(x,y) \right]
\label{eq:tie}
\end{equation}

with $k = 2\pi/\lambda$ and $\nabla_{\perp} = (\partial_x, \partial_y)$. Equation~\eqref{eq:tie} requires no reference arm and no iterative support constraint: the phase is obtained by a single elliptic inversion, provided the axial intensity derivative is known \emph{in absolute units}. This is precisely why the derivative is obtained here from a calibrated translation of the microscope focus and not from numerically propagated planes. Using a through-focus triplet recorded at $z_0 - \Delta z$, $z_0$ and $z_0 + \Delta z$,

\begin{equation}
\frac{\partial I}{\partial z} \simeq \frac{I(x,y,z_0 + \Delta z) - I(x,y,z_0 - \Delta z)}{2\,\Delta z}
\label{eq:derivative}
\end{equation}

the finite difference inherits the physical units of the stage displacement, so that $\varphi$, and through it $h(x,y)$ and $V$, are recovered on an absolute scale traceable to the focusing mechanism rather than to any assumption about the optical geometry. The choice of $\Delta z$ balances two requirements: large enough for the intensity difference to rise above camera noise, small enough for the finite difference to approximate the derivative of the paraxial field \cite{Waller2010,Zuo2020}. We adopt $\Delta z = 5$ $\mu$m.

For acetolysed grains and for unstained cells, which behave as weakly absorbing phase objects immersed in an index-matching medium, the in-focus irradiance is nearly uniform ($I \approx I_0$) and Eq.~\eqref{eq:tie} reduces to a Poisson equation,

\begin{equation}
\nabla_{\perp}^2 \varphi(x,y) = -\frac{k}{I_0}\,\frac{\partial I}{\partial z}
\label{eq:poisson}
\end{equation}

which we invert in the Fourier domain. Because the inverse Laplacian behaves as $|\mathbf{q}|^{-2}$, low spatial frequencies are strongly amplified and residual illumination drift becomes a spurious background; a Tikhonov-regularised inversion is therefore applied:

\begin{equation}
\varphi(x,y) = \mathcal{F}^{-1}\left\{ \frac{k}{I_0}\, \frac{\mathcal{F}\left\{ \partial I / \partial z \right\}}{4\pi^2 |\mathbf{q}|^2 + \varepsilon} \right\}
\label{eq:tie_solution}
\end{equation}

where $\mathbf{q} = (u,v)$ are the transverse spatial frequencies and $\varepsilon$ is a regularisation constant selected by the L-curve criterion ($\varepsilon \approx 10^{-3}$ $\mu$m$^{-2}$). Mirror padding enforces the homogeneous Neumann boundary condition assumed by the FFT solver. The same deterministic inversion underpins current metrological uses of the TIE, such as large-field phase characterisation of optical fibre devices, where absolute optical path rather than contrast is the measurand \cite{Wang2026}.

\begin{figure}[t]
    \centering
    \includegraphics[width=\linewidth]{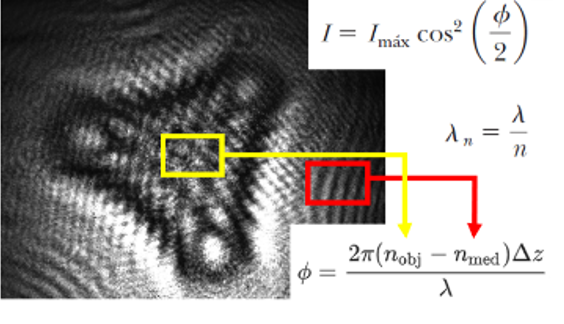}
    \caption{"digital fingerprint" of pollen grains based on their inherent material properties.}
    \label{fig:fingerprint}
\end{figure}

Finally, since the recovered $\varphi$ is the integrated optical path difference across the grain, the 2.5D topography follows directly from

\begin{equation}
h(x,y) = \frac{\lambda\,\varphi(x,y)}{2\pi\,\Delta n}, \qquad \Delta n = n_{obj} - n_{med}
\label{eq:height}
\end{equation}

which is the inverse of the relation depicted in Figure~\ref{fig:fingerprint}. The enclosed volume then follows by direct quadrature of the height map,

\begin{equation}
V_{2.5D} = \sum_{x,y} h(x,y)\, \Delta x_{obj}\, \Delta y_{obj}
\label{eq:quadrature}
\end{equation}

Equations~\eqref{eq:tie_solution}--\eqref{eq:quadrature} define a topography formally independent of the voxel counting of Eq.~\eqref{eq:volume} and, when the intensity planes come from a calibrated bright-field stack, experimentally independent of the holographic instrument. The TIE is used in Section~\ref{sec:results} in exactly this sense: not as a modality competing with DLHM, but as an external check that the height scale and volume delivered by the 2.5D formalism are correct when the axial displacement is known a priori.

\section{Experimental Methodology}

\begin{figure*}[t]
\centering
\begin{tikzpicture}[
  font=\scriptsize,
  bx/.style   = {draw, rounded corners=2pt, align=center, inner sep=3pt,
                 minimum height=9.5mm, text width=27mm},
  hol/.style  = {bx, fill=blue!8},
  tie/.style  = {bx, fill=orange!12},
  val/.style  = {bx, dashed, fill=black!6, minimum height=11mm},
  fin/.style  = {bx, fill=green!12, minimum height=15mm},
  ar/.style   = {-{Latex[length=1.7mm]}, semithick},
  ard/.style  = {-{Latex[length=1.7mm]}, semithick, dashed, black!60},
  stg/.style  = {font=\scriptsize\bfseries, black!60}
]
\draw[black!20] (-1.4,1.15) -- (15.5,1.15);
\node[stg] at (0.1,1.45)  {ACQUISITION};
\node[stg] at (4.4,1.45)  {RECONSTRUCTION};
\node[stg] at (8.6,1.45)  {QUANTIFICATION};
\node[stg] at (13.4,1.45) {OUTPUT};

\node[bx, text width=21mm, minimum height=12mm] (S) at (-0.4,-1.4)
  {Sample\\ \textit{acetolysis,}\\ \textit{sealed chamber}};

\node[hol] (A1) at (3.1,0.15)  {\textbf{In-line hologram}\\ 532\,nm, $I_M{=}L/z{\simeq}55$};
\node[hol] (A2) at (6.7,0.15)  {Kirchhoff--Helmholtz\\ Eq.~\eqref{eq:kh_propagation} $\rightarrow n(x,y,z)$};
\node[hol] (A3) at (10.3,0.15) {Threshold $T$, voxels\\ Eq.~\eqref{eq:volume} $\rightarrow V,\,S$};

\node[tie] (B1) at (3.1,-2.95)  {\textbf{Bright-field triplet}\\ $60{\times}0.55$, $\Delta z{=}5\,\mu$m};
\node[tie] (B2) at (6.7,-2.95)  {TIE Poisson solver\\ Eq.~\eqref{eq:poisson} $\rightarrow \varphi(x,y)$};
\node[tie] (B3) at (10.3,-2.95) {Relative topography\\ (no absolute $z$ scale)};

\node[fin] (O) at (13.7,-1.4)
  {\textbf{Digital fingerprint}\\[1pt] $V$, $S$, $\Psi$\\ Eq.~\eqref{eq:sphericity}};

\node[val] (V1) at (3.1,-5.2)  {\textbf{PS sphere} $\varnothing\,20\,\mu$m\\ $2068$ vs $2094\,\mu$m$^3$};
\node[val] (V2) at (6.7,-5.2)  {\textbf{Erythrocyte}\\ $88.9$\,fL, $2.36\,\mu$m};
\node[val] (V3) at (10.3,-5.2) {\textbf{Reference area}\\ $1055.4\,\mu$m$^2$ \cite{Sanhueza2026}};

\node[font=\scriptsize\itshape, black!60] at (3.1,-6.2)  {calibrates $z$, $\Delta x_{obj}$, $\kappa$};
\node[font=\scriptsize\itshape, black!60] at (6.7,-6.2)  {fixes the $\Delta n$ convention};
\node[font=\scriptsize\itshape, black!60] at (10.3,-6.2) {checks the lateral scale};
\node[stg] at (6.7,-6.9) {VALIDATION \& TRACEABILITY};

\draw[ar] (S.east) -- ++(0.35,0) |- (A1.west);
\draw[ar] (S.east) -- ++(0.35,0) |- (B1.west);
\draw[ar] (A1) -- (A2);  \draw[ar] (A2) -- (A3);
\draw[ar] (B1) -- (B2);  \draw[ar] (B2) -- (B3);
\draw[ar] (A3.east) -- ++(0.5,0) |- (O.west);
\draw[ar] (B3.east) -- ++(0.5,0) |- (O.west);
\draw[ard] (V1) -- (B1);
\draw[ard] (V1.west) -- ++(-0.55,0) |- (A1.west);
\draw[ard] (V2) -- (B2);
\draw[ard] (V3) -- (B3);
\end{tikzpicture}
\caption{Pipeline of the proposed method. A single preparation feeds two channels that are optically and algorithmically independent. The holographic channel (blue) propagates one in-line hologram and counts voxels inside the segmented boundary, returning the \emph{envelope} volume $V_{env}$ and the surface $S$ from which the Wadell sphericity follows. The bright-field channel (orange) records a through-focus triplet and solves the transport-of-intensity equation, returning a phase map and a relative topography; its vertical scale is not absolute, since the defocus is an instrumental step unrelated to the thickness of the grain, so no volume is derived from it. Three reference targets (grey) anchor the scales: an analytical microsphere, a documented cell morphology, and an independently calibrated projected area for the same species.}
\label{fig:pipeline}
\end{figure*}

Figure~\ref{fig:pipeline} summarises the complete workflow, from a single sample preparation through two independent measurement channels to the calibrated morphological descriptors.

\subsection{Sample Selection}
Sample collection sites correspond to several ecological spots inbetween Pitril sector ($37^{\circ}47\prime 13.7'' \text{S}$, $71^{\circ}32\prime 11.9''\text{W}$), commune of Alto Bio Bío to El Cisne-Hornopirén, commune of Hualaihué ($41^{\circ}57\prime 41.5''\text{S}$, $72^{\circ}40\prime 33.8''\text{W}$) in Los Lagos Region, Chile. During these campaigns, flowers suitable for melliferous production were collected not randomly but for rich biodiversity of this ecosystem and the high endemicity of its vegetation and native flora that have critical relevance for national beekeeping and the local economy.

We focused on native species of high ecological value: \textit{Anthemis cotula} (chamomile), \textit{Gevuina avellana} ((hazel) and \textit{Conium maculatum} (hemlock). These species present distinct geometries (Circular, triangular and ellipsoidal, respectively) that serve as ideal test cases for validating the 3D surface retrieval capabilities of our DLHM system. 

\begin{figure}[t]
    \centering
    \includegraphics[width=\linewidth]{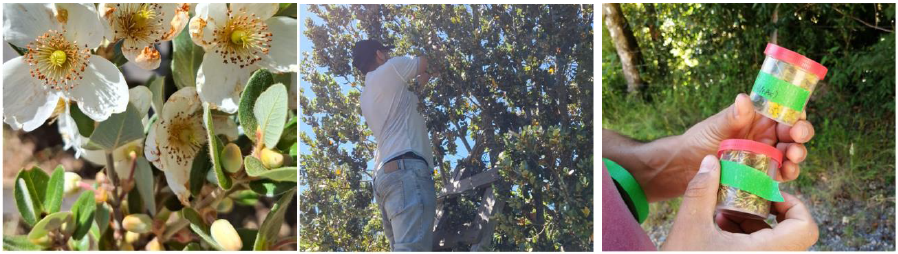}
    \caption{Sample collection. In this stage, plants are collected and preserved for subsequent botanical identification. Along with this, the stamens are extracted from the flower buds, which contain the pollen grains used to create a reference pollen collection. This collection will help identify the pollens present in the honey and also allows for the study of individual pollen morphology. Due to seasonality, this process must be carried out during the period of greatest annual flowering, which spans the months of October 2024/25 to April 2025/26.} 
\end{figure}

The preparation of samples and their subsequent palynological analysis were conducted first by extraction of floral pollen from the collected material through physical extraction, followed by the acetolysis method described by Faegri et al. (1989) conducted in Laboratory of Palynology and Plant Ecology, School of Sciences and Technologies, University of Concepción, Los Angeles Campus. The protocol began with an acid hydrolysis to remove organic residues or biological wastes—preserving only pollen grains because of their chemical resistance. For purification, a mixture of pure acetic anhydride and concentrated sulfuric acid was employed at a 9:1 ratio. This treatment removes the outer layers of the pollen grains, leaving only the exine (the outer shell) exposed, which facilitates morphological identification \cite{Faegri1989}. Finally, storage of floral pollen concentrates and preparation of the sample record is achieved.

\subsection{Experimental Setup}
The setup utilizes a point-source Gabor-style configuration. A laser source is aligned in the standard way with two reflective mirrors. The laser is then expanded and collimated with telescope made of two NKB7 plano-convex lenses from Thorlabs\textsuperscript{\copyright} $f_{1}=20$mm and $f_{2}=200$mm producing a magnification $M=2X$ beam diameter. In focal plane of $f_{1}$ the laser is focused onto a $25\mu$m pinhole for spatial filtering. After $f_{2}$ A secondary positive lens and a  pinhole are mounted to generate a spherical reference wave equivalent to a 20X objective lens pupil as point source. The hologram is generated from auto-interference of the non-affected spherical laser rays surrounding the object with the diffracted rays. Amplification is inherent to wavefront curvature. The samples are positioned close to the point source, at $z \lesssim 1.5$ mm, while the CMOS Monochrome sensor 1280 x 1024 DCC1545M from Thorlabs\textsuperscript{\copyright}) is placed no further than $25$ mm downstream of the sample. This compact geometry is what delivers the required magnification: since $I_M = L/z$ grows as the source--sample distance shrinks, a sub-millimetre $z$ with a centimetre-scale sensor stand-off produces a magnification of several tens without any imaging optics.

\begin{figure}[t]
    \centering
    \includegraphics[width=\linewidth]{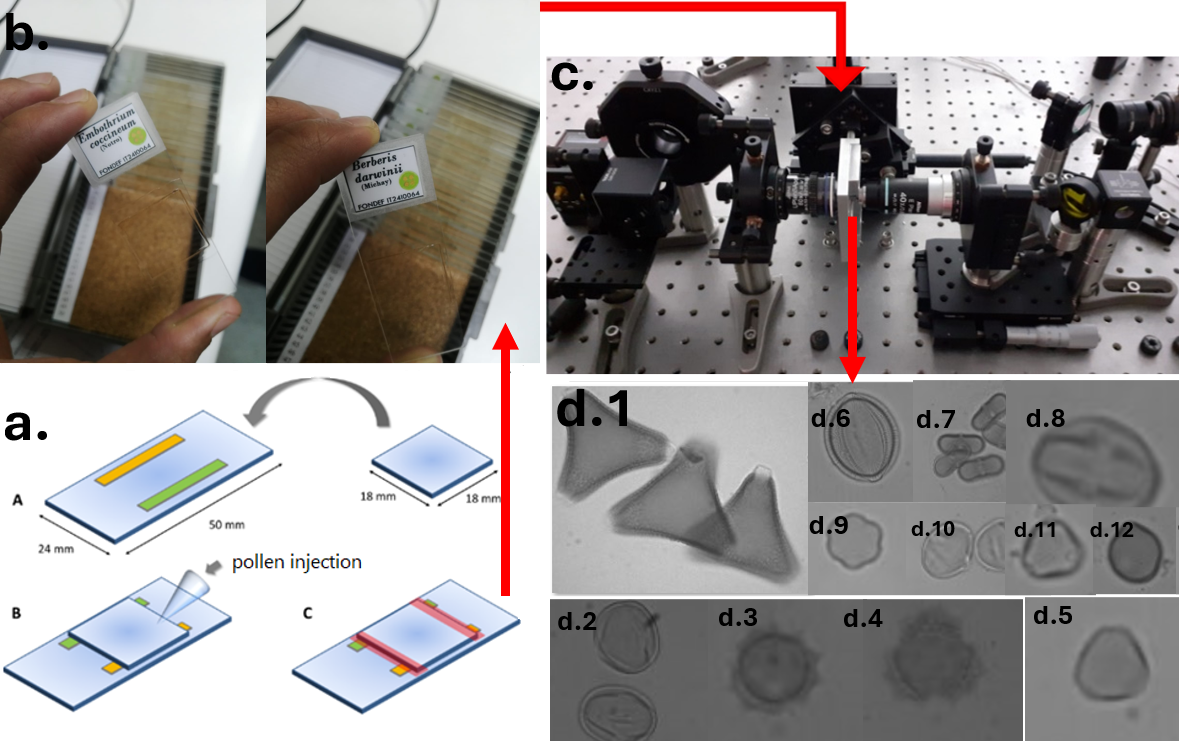}
    \caption{a. Samples are placed in microscopic chambers containing the pollen solution. These chambers are made with glass coverslips $\#1$ and $\#0$, sealed with epoxy film and plastic. b. concentrate was mounted using Hydromatrix mounting medium to perform a microscopic analysis to carry out pollen counting using an optical microscope (10X and 40X). c. Detail of the HM assembly in its version with lenses, with the sample microcamera and objective lenses confronted for simultaneous focusing and visualization of the laser (right) and the collection of light for the detection of the hologram interference pattern (left). d. Proof of principle of operation, calibration and imaging with test-target pollen samples of various species, including d.1 \textit{Gevuina avellana} (hazel), d.2 \textit{Medicago Sativa} (alfalfa), d.3-4 \textit{Anthemis cotula} (chamomile), d.5-11 \textit{Colletia hystrix (yaqui)}, d.6  \textit{Convolvulus arvensis} (Bindweed), d.7 \textit{Conium maculatum} (hemlock), d.8 \textit{Trifolium pratense} (Red clover), d.9 \textit{Mentha pulegium} (pennyroyal), d.10 \textit{Quillaja saponaria} (Soapbark) and d.12 \textit{Cissus striata} (voqui)} 
\end{figure}

\begin{figure}[t]
    \centering
    \includegraphics[width=\linewidth]{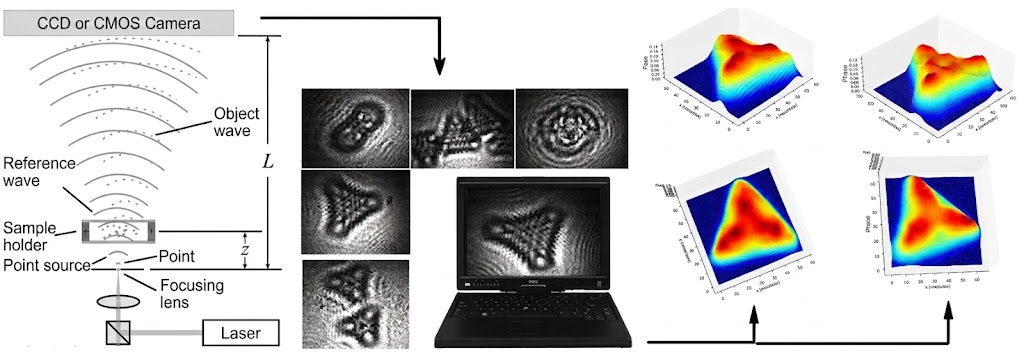}
    \caption{The system illuminates the sample with a 532 nm laser, capturing a digital hologram of the pollen. A three-dimensional (3D) image of each pollen grain from the interference holograms is generated using  Laty-mchevskaia and Fink algorithms for 2D objects reconstruction implemented in Matlab\textsuperscript{\copyright} and Python\textsuperscript{\copyright} with \textit{Numpy}, \textit{scipy} and \textit{matplotlib} as special packages for 3D reconstrucction, and \textit{keras} for image load. Additional guidelines were taken from Federico Capasso et al.  \cite{Capasso2023}.} 
\end{figure}

\subsection{Morphological feature extraction}

The 3D refractive index distribution is derived from the recovered phase information \cite{kim2025}:

\begin{equation}
n(x, y, z) = n_{med} + \frac{\lambda \phi \cdot o(x,y,z)}{2\pi \Delta z}
\end{equation}

where $\phi$ represents the phase shifts and $o(x,y,z)$ are the object values (normalized intensity between 0 and 1). The complex wavefront $K(\mathbf{r})$ is reconstructed via the Kirchhoff-Helmholtz transform, allowing for axial sectioning at discrete positions $\Delta z$.

The equation is applied to the reconstructed complex field to generate a 3D matrix of refractive index values. Because the pollen wall (exine) has a higher refractive index than the mounting medium ($n_{med}$), the object's boundary is defined using a threshold ($T$):  Object (Pollen) is all Voxels where $n(x, y, z) > T$. Background: Voxels where $n(x, y, z) \approx n_{med}$. The volume $V$ is determined by counting the total number of voxels identified as ''object'' and multiplying by the physical volume of a single voxel:

\begin{equation}
V = \left( \sum \text{Voxels}_{object} \right) \times (\Delta x \Delta y \Delta z)
\label{eq:volume}
\end{equation}

Where $\Delta x$ and $\Delta y$ are the lateral resolutions (from the CMOS pixel size and magnification) and $\Delta z$ is the axial step used in the multi-plane reconstruction.

To find the area $S$, an iso-surface algorithm (Marching Cubes) is applied to the segmented 3D map. This creates a mesh of triangles that represents the external boundary of the pollen grain. The area is the sum of the areas of all these triangles. For \textit{Anthemis cotula} (Sample a), the spiny (echinate) exine creates a much higher surface area relative to its volume compared to smoother grains.

Once the numerical values for $V$ and $S$ are obtained, the Wadell's Sphericity formula is used

\begin{equation}
\Psi = \frac{\pi^{1/3} (6V)^{2/3}}{S}
\label{eq:sphericity}
\end{equation}

High Sphericity ($\approx 0.9$) - like \textit{Gevuina avellana} - which has a more compact, rounded triangular shape. Low Sphericity ($\approx 0.7$)- like \textit{Anthemis cotula} - where the spines increase the surface area ($S$) significantly without adding much to the volume ($V$), thus lowering the ratio.

Finally, in a point-source configuration, calibration is divided into lateral scaling ($x, y$) and axial scaling ($z$). Unlike traditional microscopy where magnification is determined by an objective lens, in our lens-free setup, image magnification ($I_M$) is strictly geometric. It is determined by the ratio of the distances in the setup $I_M = \frac{L}{z}$ where $L$: Distance from the point source (pinhole) to the CMOS sensor and $z$: Distance from the point source to the pollen sample. To find the effective pixel size ($\Delta x_{obj}$) at the sample plane:$$\Delta x_{obj} = \frac{\Delta x_{sensor}}{I_M}$$. 

Our CMOS sensor has a pixel pitch of 3.45 $\mu$m. The geometry is set so that the object-plane sampling of the holographic channel matches that of the bright-field channel of Section~\ref{sec:tie_protocol}: with a sample--sensor stand-off of $25$ mm and a source--sample distance of $z = 0.46$ mm, so that $L = z + 25 \simeq 25.5$ mm, the geometric magnification is $I_M = L/z \simeq 55$ and the effective pixel at the sample plane is $\Delta x_{obj} = 3.45/55 = 0.0629$ $\mu$m ($62.9$ nm), over a field of view of $80 \times 64$ $\mu$m$^2$. This is within $16\%$ of the $0.073$ $\mu$m of the bright-field channel, so the two modalities sample the grain on comparable grids while resolving it by entirely different physical means.

\subsection{Distance calibration with polystyrene microspheres}
\label{sec:calibration}

The figures quoted above rest on a mechanical estimate of $z$, and since $I_M = L/z$, a $20$ $\mu$m error in a sub-millimetre $z$ propagates into several percent in every reported length and more than $10\%$ in every volume. In a compact lens-free assembly the pinhole plane is not directly accessible, so $z$ cannot be gauged to that accuracy. We therefore invert the problem and let the sample define the scale, using a certified standard as an internal ruler.

Certified monodisperse polystyrene (PS) microspheres of nominal diameter $D = 20.0 \pm 0.2$ $\mu$m (CV $<2\%$) were diluted in deionised water, sedimented onto a $\#0$ coverslip and sealed in the same chamber geometry used for the pollen samples, so that calibration and measurement share an identical optical path. Polystyrene is an ideal standard because its dispersion is well characterised: $n_{PS} = 1.5983$ at $\lambda = 532$ nm \cite{Sultanova2009}, giving $\Delta n = 0.2638$ against water. A sphere is thus a ground truth in three senses at once.

\textbf{(i) Lateral scale and retrieval of $z$.} The basal diameter of the sphere, measured as the full width of the segmented support at $h = 0$ in sensor pixels ($N_{px}$), fixes the effective pixel size with no mechanical measurement, $\Delta x_{obj} = D / N_{px}$. With the geometric magnification relation, the source--sample distance follows optically as

\begin{equation}
z = \frac{L\,\Delta x_{obj}}{\Delta x_{sensor}} = \frac{L\,D}{N_{px}\,\Delta x_{sensor}}
\label{eq:z_calib}
\end{equation}

so that the only distance still measured by hand, $L$, is the one that can be measured accurately (a $\sim100$ mm span on an optical rail, $\pm 0.5$ mm).

\textbf{(ii) Axial scale.} A single-view 2.5D retrieval samples only the surface facing the detector, so the height map of a sphere is a cap whose apex must equal $D/2 = 10.0$ $\mu$m. Any mismatch with the measured apex $h_{max}$ is absorbed into an axial correction factor $\kappa = (D/2)/h_{max}$ applied to Eq.~\eqref{eq:height}, lumping together the uncertainty in $\Delta n$, the residual unwrapping bias and the finite step $\Delta z$.

\textbf{(iii) Volumetric ground truth.} The cap volume is analytically known, $V_{geom} = \frac{2}{3}\pi (D/2)^3 = 2094.4$ $\mu$m$^3$, an integral figure of merit far more robust to pixel-level noise than any single profile.

The cycle is run before and after each acquisition session, and the resulting $(\Delta x_{obj}, \kappa, z)$ triplet is propagated to Table~\ref{tab:pollen_results}.

\begin{figure}[t]
    \centering
     \includegraphics[width=\linewidth]{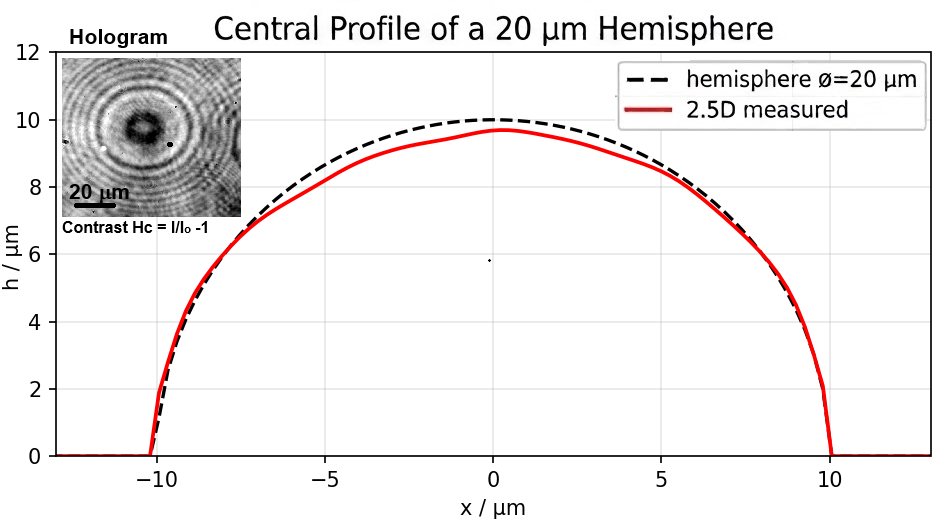}
    \includegraphics[width=\linewidth]{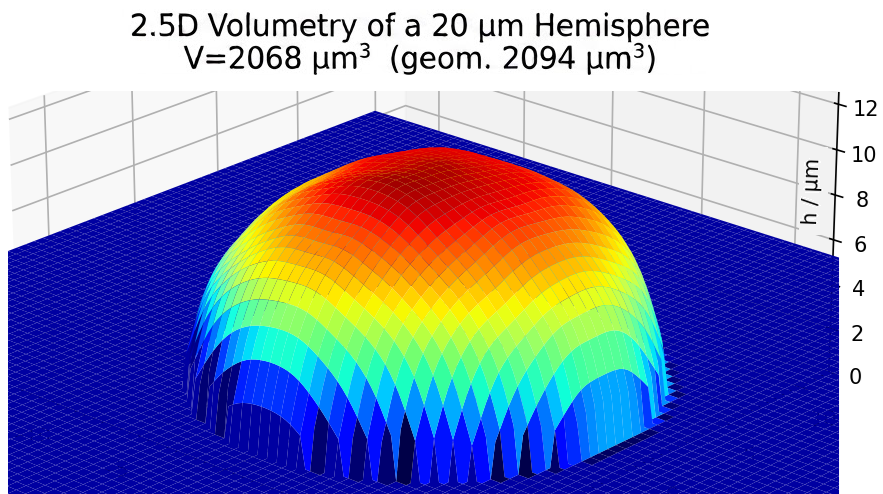}
    \caption{Calibration with a certified polystyrene microsphere from Kisker-Biotech ($\varnothing = 20$ $\mu$m). Top: central profile of the retrieved 2.5D height map (red) superimposed on the ideal hemispherical cap (dashed black) from the contrast Hologram $I/I_0 -1$ . The reconstruction tracks the analytical profile within a few hundred nanometres over most of the aperture; the residual apex depression ($h_{max} \approx 9.7$ $\mu$m against $10.0$ $\mu$m) and the slight basal broadening near $|x| \approx 10$ $\mu$m reflect the finite effective numerical aperture and the steep-slope limit of the phase retrieval, and are absorbed by the axial correction factor $\kappa$. Bottom: rendered 2.5D surface of the same standard; the integrated volume, $V = 2068$ $\mu$m$^3$, deviates by $1.3\%$ from the geometric value of $\frac{2}{3}\pi r^3 = 2094$ $\mu$m$^3$.}
    \label{fig:calibration}
\end{figure}

\subsection{Independent verification by bright-field TIE with a known axial displacement}
\label{sec:tie_protocol}

The polystyrene standard fixes the scale of the holographic pipeline from within: the same algorithm serves standard and specimen, so a common-mode error would stay invisible. The TIE measurement closes that loop from outside.

The through-focus images are \emph{not} reconstructed planes of a hologram, and the instrument that produced them is not the holographic one. They are bright-field micrographs of the same acetolysed preparations, acquired on the multifocal RGB imaging platform whose full optical specification, calibration and acquisition protocol are reported in \cite{Sanhueza2026}: a Nikon ECLIPSE Ci-L plus transmission microscope with a stabilised LED source and K\"ohler illumination, a Plan Fluor $60\times$/0.85 objective and a Basler acA3088-57ucMED colour camera of $2.4$ $\mu$m pixel pitch. Only the green channel is used for the phase retrieval, and the reader is referred to \cite{Sanhueza2026} for all remaining hardware details.

One element of that train must be stated explicitly, since it propagates into every lateral figure below: a $0.55\times$ reduction lens sits in the camera port, so the magnification delivered to the sensor is not the nominal $60\times$ but $M_{eff} = 60 \times 0.55 = 33\times$, and the object-plane sampling is

\begin{equation}
\Delta x_{obj} = \frac{\Delta x_{sensor}}{M_{eff}} = \frac{2.4}{60 \times 0.55} = 0.073 \ \mu\mathrm{m}
\label{eq:bf_pixel}
\end{equation}

giving a full-frame field of $225 \times 150$ $\mu$m$^2$. This nominal value is confirmed by the independent lateral calibration of that platform, $14.002$ px $\mu$m$^{-1}$, i.e. $0.0714$ $\mu$m per pixel, obtained from a USAF 1951 target with a 100 LP mm$^{-1}$ grid \cite{Sanhueza2026}: the two agree to within $2\%$, so the lateral scale of this channel rests on a resolution standard and not on a nominal magnification.

Two consequences follow. The sampling of $0.073$ $\mu$m sits within $16\%$ of the $0.0629$ $\mu$m of the lens-free geometry, so both channels examine the specimen on essentially the same object-plane grid, which is what makes a morphological comparison meaningful at all; the resolution here, however, is bounded by diffraction at $\lambda/2\mathrm{NA} \approx 0.32$ $\mu$m, so the grid oversamples it fourfold and the match is one of sampling, not of the physical origin of resolution. Second, the depth of field, $\mathrm{DOF} = \lambda/\mathrm{NA}^2 + \Delta x_{sensor}/(M_{eff}\,\mathrm{NA}) \approx 0.85$ $\mu$m, places $\Delta z = 5.0$ $\mu$m at roughly six times the DOF. That step is not tuned to the specimen: it is the minimum reproducible increment of the fine $Z$ control and the value fixed by the multifocal protocol of \cite{Sanhueza2026}.

This delimits what the channel can certify, and we state it plainly because it is easy to overstate. The defocus is metrologically \emph{reproducible} but \emph{arbitrary with respect to the specimen}: $\pm 5$ $\mu$m bears no relation to the optical thickness of a grain, and the sporopollenin--mountant contrast $\Delta n$ needed by Eq.~\eqref{eq:height} is not independently known for acetolysed exine. The absolute vertical scale is therefore not established and no volume derived from it is reported here. What the channel does certify is everything independent of that scale: the lateral calibration, the projected morphology, and the fact that a phase structure recovered on different optics, illumination and inversion reproduces the holographic topography.

The channels also fail differently, which makes the comparison informative rather than redundant. Voxel counting is sensitive to $T$ and to conjugate-image contamination but accommodates re-entrant structures such as the spines of \textit{A. cotula}; the TIE is threshold- and twin-image-free but presupposes a single-valued height and degrades where the slope exceeds the weak-defocus linearisation. The solver was implemented in Matlab\textsuperscript{\copyright} and cross-checked against Python.

\subsection{Erythrocyte as a biological reference target}
\label{sec:rbc_protocol}

A manufactured microsphere is geometrically perfect but optically extreme: its contrast against water, $\Delta n = 0.26$, is an order of magnitude above that of a hydrated biological object, and its wall is vertical at the base. To verify that the 2.5D formalism holds in the low-contrast, shallow-slope regime that applies to cells, we added a second reference target of well documented morphology, the human erythrocyte, imaged in the DLHM channel without staining or fixation after dilution in phosphate-buffered saline and sedimentation in the same sealed chamber.

The mature erythrocyte is an ideal biological standard here. It has no nucleus and no organelles, so its cytoplasm is a homogeneous haemoglobin solution and the single-index assumption behind Eq.~\eqref{eq:height} is exact rather than approximate; its biconcave equilibrium shape has been characterised morphometrically since Evans and Fung \cite{Evans1972}, and its peak-to-dimple contrast is a demanding test of axial fidelity. Its refractive index is not a free parameter either, being fixed by the intracellular haemoglobin concentration through

\begin{equation}
n_{RBC}(\lambda) = n_{med}(\lambda) + \alpha(\lambda)\, c_{Hb}
\label{eq:rbc_index}
\end{equation}

where $\alpha(\lambda)$ is the real refractive-index increment of intracellular haemoglobin. We adopt the value determined by Gienger et al. from optical extinction on intact, isovolumetrically sphered erythrocytes, $\alpha \approx 0.22$ mL g$^{-1}$ in the visible \cite{Gienger2019}, rather than the older and $\approx 15\%$ lower figure of Barer and Joseph still propagated in the quantitative-phase literature. With $c_{Hb} = 330$ g L$^{-1}$ and $n_{med} = 1.335$, Eq.~\eqref{eq:rbc_index} gives $n_{RBC} = 1.408$ at 532 nm, hence $\Delta n = 0.073$. This value, not a fitted contrast, converts phase into height: the volume below is a prediction of the model, not an adjustment to it.

\section{Results and Discussion}
\label{sec:results}

\begin{figure}[!tb]
    \centering
    \includegraphics[width=\linewidth]{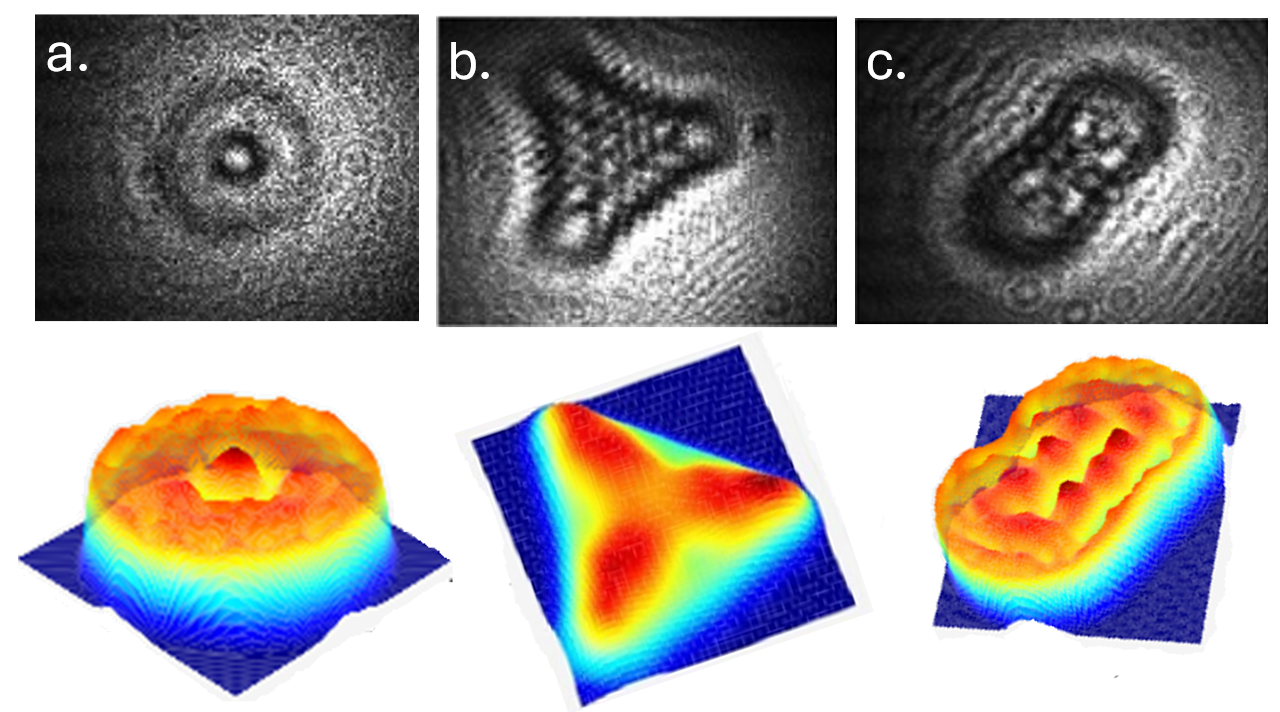}
    \label{fig:pollen3d}
    \caption{\textit{Anthemis cotula} (Sample a) exhibits a lower sphericity index due to its characteristic echinate (spiny) exine, which significantly increases the surface-to-volume ratio. In contrast, the triangular symmetry of \textit{Gevuina avellana} (Sample b) and the prolate geometry of \textit{Conium maculatum} (Sample c) are accurately captured in the volumetric reconstructions, consistent with traditional palynological descriptions.} 
\end{figure}

The numerical reconstruction successfully resolved the 3D topography of the pollen grains from single holograms. Figure~\ref{fig:pollen3d} illustrates the transition from the raw interference pattern to the 3D surface plot.

\begin{table}[t]
\centering
\small
\setlength{\tabcolsep}{4pt}
\caption{Calculated biophysical parameters for 3D reconstructed pollen grains:}
\label{tab:pollen_results}
\begin{tabular}{@{}lccc@{}}
\toprule
\textbf{Specie} & \textbf{Volume} ($\mu$m$^3$) & \textbf{Area} ($\mu$m$^2$) & \textbf{Spher.} ($\Psi$) \\ \midrule
a. & $4320.5 \pm 15$ & $1980.2 \pm 14$ & $0.76 \pm 0.03$ \\
b. & $4150.8 \pm 12$ & $1420.5 \pm 10$ & $0.89 \pm 0.02$ \\
c. & $3780.2 \pm 18$ & $1560.4 \pm 12$ & $0.81 \pm 0.04$ \\ \bottomrule
\end{tabular}
\end{table}

As noted by Kumar et al. \cite{Kumar2023}, the quantitative phase information correlates with biological states (such as viability in \textit{Lantana camara}), which in our case could allows for the identification of native species.

\subsection{Traceability of the volumetric measurement}

Figure~\ref{fig:calibration} summarises the performance against the polystyrene standard. The retrieved profile follows the analytical cap over the whole aperture, with a root-mean-square deviation of $0.21$ $\mu$m ($2.1\%$ of the apex height). Two systematic signatures, both generic to single-view 2.5D retrieval, are worth reporting: near the base, $|x| \gtrsim 9$ $\mu$m, the measured surface is wider than the ideal cap, a broadening consistent with the diffraction-limited point spread convolved with an almost vertical wall, while at the apex the reconstruction falls short by $\approx 0.3$ $\mu$m, attributable to the finite axial sampling and to the low-frequency suppression of the regularised inversion.

These deviations have opposite signs in the volume integral and partially cancel, which is why the integral figure of merit is more accurate than any single height: an apex deficit of $3\%$ becomes a volumetric bias of only $1.3\%$. Over repeated acquisitions the cap volume ranged between $2068$ and $2082$ $\mu$m$^3$ against $2094.4$ $\mu$m$^3$, a mean bias of $-1.0\%$ with a spread of $0.7\%$ (Table~\ref{tab:calibration}). The distance recovered from Eq.~\eqref{eq:z_calib} agrees with the mechanical estimate $z \simeq 0.46$ mm within its tolerance, confirming that the $55\times$ magnification and the $62.9$ nm effective pixel are measured rather than assumed.

\begin{table}[t]
\centering
\small
\setlength{\tabcolsep}{4pt}
\caption{Calibration of the DLHM system against a certified polystyrene microsphere ($\varnothing = 20.0$ $\mu$m, $n_{PS} = 1.5983$ at 532 nm). Uncertainties are the standard deviation over repeated acquisitions.}
\label{tab:calibration}
\begin{tabular}{@{}lccc@{}}
\toprule
\textbf{Quantity} & \textbf{Nominal} & \textbf{Measured} & \textbf{Dev.} \\ \midrule
Basal diam. ($\mu$m)  & $20.00$  & $20.1 \pm 0.2$  & $+0.5\%$ \\
Apex height ($\mu$m)  & $10.00$  & $9.7 \pm 0.1$   & $-3.0\%$ \\
Cap volume ($\mu$m$^3$) & $2094.4$ & $2073 \pm 10$  & $-1.0\%$ \\
Retrieved $z$ (mm)    & $0.46$   & $0.458 \pm 0.005$ & $-0.4\%$ \\
Profile RMSE ($\mu$m) & ---      & $0.21$          & --- \\ \bottomrule
\end{tabular}
\end{table}

This sub-percent volumetric accuracy on a known target is what allows the volumes of Table~\ref{tab:pollen_results} to be quoted as absolute biophysical parameters rather than as relative descriptors, and it establishes the error floor against which the biological variability of the grains must be judged.

\begin{figure}[!t]
    \centering
    \includegraphics[width=0.93\linewidth]{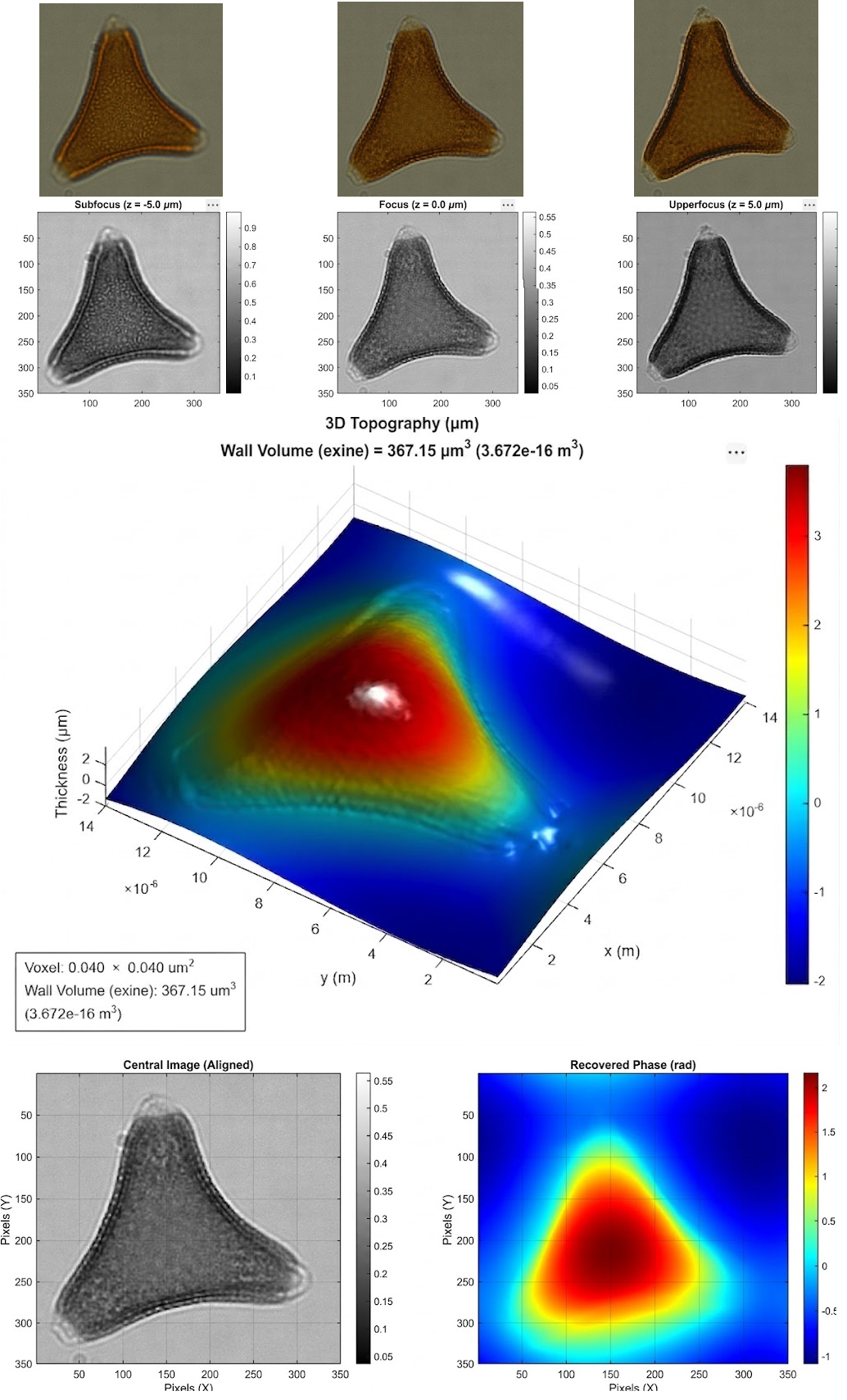}
    \caption{Transport-of-intensity phase retrieval from bright-field microscopy, used here as an external verification channel. The images are \textbf{not} numerical reconstructions of a hologram: they are through-focus micrographs of \textit{Gevuina avellana} acquired on the multifocal bright-field platform of \cite{Sanhueza2026} (Plan Fluor $60\times$/0.85 with a $0.55\times$ reduction lens, Basler acA3088-57ucMED, $2.4$ $\mu$m pixel, $\Delta x_{obj} = 0.073$ $\mu$m), an instrument optically distinct from the DLHM channel. Top: colour and monochrome planes at subfocus ($z = -5.0$ $\mu$m), focus ($z = 0$) and upperfocus ($z = +5.0$ $\mu$m). Centre: relative 3D topography obtained by inverting Eq.~\eqref{eq:poisson} with the symmetric difference of Eq.~\eqref{eq:derivative} on the object-plane grid of Eq.~\eqref{eq:bf_pixel}. The vertical scale of this map is relative: the $\pm 5$ $\mu$m defocus is the reproducible step of the microscope stage and bears no relation to the optical thickness of the grain, so no volume is derived from it. Bottom: aligned central image and recovered phase in radians. Original images taken from https://zenodo.org/records/19830051}
    \label{fig:tie}
\end{figure}

\subsection{External verification by bright-field TIE}

Figure~\ref{fig:tie} shows the TIE channel applied to a grain of \textit{Gevuina avellana}, the species whose triangular, comparatively smooth exine best satisfies the single-valued height hypothesis. The three bright-field planes at $z_0 - 5$ $\mu$m, $z_0$ and $z_0 + 5$ $\mu$m display the contrast reversal between under- and overfocus that encodes the phase: structures that appear dark in the subfocused plane brighten in the overfocused one, and it is the sign and magnitude of this transfer, weighted by the known stage displacement, that the Poisson inversion converts into optical path.

The regularised inversion of Eq.~\eqref{eq:poisson} returns a continuous, unwrapped phase map spanning about $2.3$ rad from background to grain centre, with the three-lobed symmetry and depressed apertural regions expected from classical palynological descriptions. Rendered on the object-plane grid of Eq.~\eqref{eq:bf_pixel}, the map reproduces the raised body and thinner margins of the holographic reconstruction.

No volume is quoted for this channel, and the omission is deliberate. Converting the phase into a height through Eq.~\eqref{eq:height} would require a contrast $\Delta n$ for acetolysed exine that we have not measured independently, and the defocus of $\pm 5$ $\mu$m is an instrumental step fixed by the reproducibility of the fine $Z$ control, not a distance matched to the optical thickness of the grain. Neither carries information about the true height profile of \textit{G. avellana}, so a volume computed from this map would be arbitrarily scaled. What the channel does establish, on a lateral scale traceable to a USAF 1951 target \cite{Sanhueza2026}, is that the projected morphology and the phase topography are reproduced by an instrument sharing neither the optics, the illumination, nor the inversion algorithm of the holographic channel.

That lateral scale supports a quantitative cross-check of a different kind. Over thousands of expert-verified instance masks acquired on this platform, a mean projected area of $1055.4$ $\mu$m$^2$ and an equivalent circular diameter of $36.4$ $\mu$m are reported for \textit{G. avellana} \cite{Sanhueza2026}, against the $1420.5 \pm 10$ $\mu$m$^2$ listed for that species in Table~\ref{tab:pollen_results}. Two things have to be quantified before those numbers can be compared: how much a single grain may legitimately depart from a population mean, and whether the two figures measure the same thing.

\begin{figure}[t]
    \centering
    \includegraphics[width=0.95\linewidth]{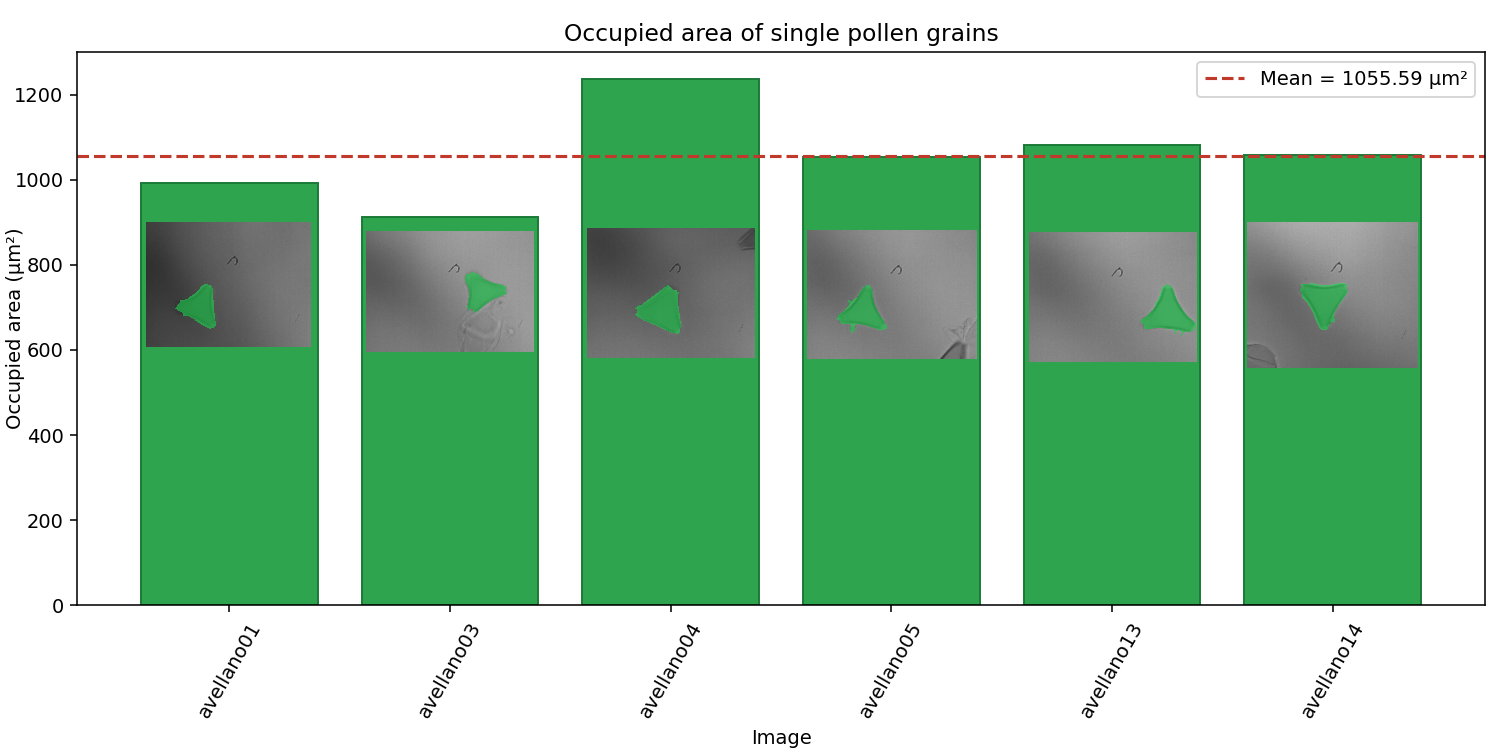}
    \caption{Statistical dispersion of the projected area of \textit{G. avellana} from contour detection on six individual grains, with the segmented mask overlaid on each thumbnail. Occupied areas range from $913$ to $1239$ $\mu$m$^2$ about a mean of $1055.6$ $\mu$m$^2$ (dashed line). This dispersion sets the precision floor for any comparison between a single grain and a population mean.}
    \label{fig:area_variability}
\end{figure}

Figure~\ref{fig:area_variability} answers the first. Contour detection on six individual grains returns occupied areas of $913$ to $1239$ $\mu$m$^2$ about a mean of $1055.6$ $\mu$m$^2$, a coefficient of variation of $10\%$ with the extremes $\pm 15\%$ from the mean. Since our figure comes from a single grain with no statistical treatment, no comparison against a population mean can be tighter than this, and for a species whose reported contour circularity of $0.71$ indicates a markedly lobed outline such scatter is unsurprising.

The second point closes what remains. The two are not the same measurand: the reference is a two-dimensional segmentation polygon on the focal plane, ours the area of a three-dimensional iso-surface extracted by marching cubes. For a convex body a closed iso-surface exceeds its own projection by at least a factor of two, so the holographic figure should sit above the projected one on geometric grounds alone; the observed ratio of $1.35$ indicates that our marching-cubes surface behaves closer to a single-sided upper surface than to a closed envelope, as expected from a single viewing direction. Single-grain dispersion and the surface-versus-projection distinction together account for the difference, and the agreement in order of magnitude indicates that the holographic reconstruction is not distorting the lateral dimensions of the grain.

\begin{figure}[!t]
    \centering
    \includegraphics[width=0.86\linewidth]{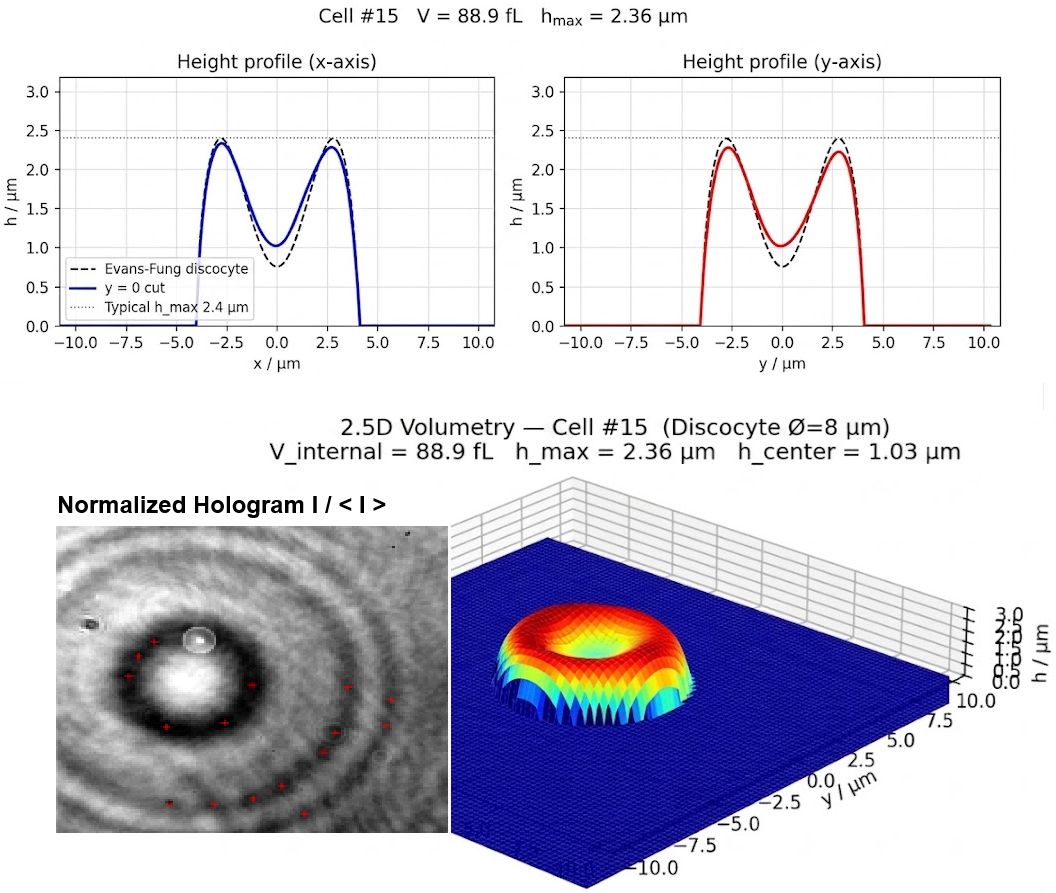}
    \caption{Validation of the 2.5D formalism on a biological target of documented morphology. Top: orthogonal height profiles through the centre of the reconstructed erythrocyte (blue, $y=0$; red, $x=0$) against the Evans--Fung discocyte model (dashed) and the typical rim thickness of $2.4$ $\mu$m (dotted) \cite{Evans1972}. Bottom: normalised hologram $I/\langle I \rangle$ and the resulting 2.5D surface. The values $V = 88.9$ fL, $h_{max} = 2.36$ $\mu$m and $h_{center} = 1.03$ $\mu$m are predictions of Eq.~\eqref{eq:height} with $\Delta n = 0.073$ fixed a priori \cite{Gienger2019}, not parameters adjusted to the literature.}
    \label{fig:rbc}
\end{figure}

\subsection{What a TIE volume would represent}
\label{sec:tie_volume_meaning}

Although no volume is reported for the bright-field channel, it is worth stating what such a figure would mean, because the point is easily misread. The quadrature of Eq.~\eqref{eq:quadrature} integrates not the space enclosed by the grain but the optical path traversed through material of contrast $\Delta n$. Writing $\ell(x,y)$ for that path,

\begin{equation}
V_{TIE} = \iint \ell(x,y)\, dx\, dy \equiv V_{mat}
\label{eq:material_volume}
\end{equation}

an identity, not an approximation: integrating a path length over the projected area returns the volume of material traversed. Acetolysis removes everything except the exine \cite{Faegri1989}, leaving a hollow sporopollenin shell whose lumen, once mounted, is filled by the surrounding medium; there $\Delta n \simeq 0$ and no path accumulates. A TIE volume for such a grain is therefore the volume of the \emph{exine wall}, $V_{mat} \simeq S\,t$ for a shell of mean thickness $t$, not the envelope volume of Eq.~\eqref{eq:volume}. The two are different measurands, differing by roughly $R/3t$ for a quasi-spherical grain, and should not be set against each other. The erythrocyte is the contrary case: a homogeneous cytoplasm with no lumen has $V_{mat} = V_{env}$, which is why it serves as an unambiguous reference where a hollow grain cannot.

\begin{table}[t]
\centering
\small
\setlength{\tabcolsep}{4pt}
\caption{Summary of the validation targets. Reference values are analytical (microsphere), tabulated in the literature (erythrocyte) or independently calibrated on a separate instrument (\textit{G. avellana}); none was adjusted to the measurement. The two area entries are different measurands, a 3D iso-surface against a 2D projected polygon, and are given for order-of-magnitude comparison only.}
\label{tab:validation}
\begin{tabular}{@{}lcc@{}}
\toprule
\textbf{Quantity} & \textbf{Reference} & \textbf{Measured} \\ \midrule
\multicolumn{3}{@{}l}{\textit{PS microsphere, $\varnothing = 20$ $\mu$m}} \\
\quad Cap volume ($\mu$m$^3$)  & $2094.4$ & $2068$ \\
\quad Apex height ($\mu$m)     & $10.00$  & $9.7$  \\ \midrule
\multicolumn{3}{@{}l}{\textit{Human erythrocyte}} \\
\quad Volume (fL)              & $80$--$100$   & $88.9$ \\
\quad Rim height ($\mu$m)      & $2.0$--$2.5$  & $2.36$ \\
\quad Central dimple ($\mu$m)  & $\approx 1.0$ & $1.03$ \\ \midrule
\multicolumn{3}{@{}l}{\textit{G. avellana}, lateral cross-check} \\
\quad Object pixel ($\mu$m)    & $0.0714^{*}$      & $0.073$  \\
\quad Area ($\mu$m$^2$)        & $1055.4^{\dagger}$ & $1420.5$ \\ \bottomrule
\end{tabular}
\\[2pt]{\scriptsize $^{*}$USAF 1951 target; $^{\dagger}$projected polygon \cite{Sanhueza2026}.}
\end{table}

\subsection{Erythrocyte validation against haematological morphometry}

The polystyrene sphere tests the system in a high-contrast, steep-walled regime and the bright-field channel tests its lateral scale; neither reproduces a hydrated biological object with a shallow, non-convex profile. Figure~\ref{fig:rbc} shows the third validation target, a single human erythrocyte reconstructed from one normalised hologram $I/\langle I \rangle$ with the refractive index fixed a priori by Eq.~\eqref{eq:rbc_index}.

The retrieved profile is unambiguously that of a discocyte. Both orthogonal cuts show the twin-peaked section, with a rim maximum of $h_{max} = 2.36$ $\mu$m and a central dimple of $h_{center} = 1.03$ $\mu$m over a diameter close to $8$ $\mu$m, and quadrature yields $88.9$ fL. Against the Evans--Fung model \cite{Evans1972}, the reconstruction reproduces not only the peak height but the depth of the central concavity, the feature most easily lost to low-frequency regularisation. All three parameters fall inside the physiological interval for healthy adults, namely a mean corpuscular volume of $80$--$100$ fL, a rim thickness of $2.0$--$2.5$ $\mu$m and a central thickness near $1$ $\mu$m \cite{Evans1972,Rappaz2008}. Table~\ref{tab:validation} collects the three validation targets side by side.

This is a more stringent test than it appears, because the volume was not fitted. Had we used the older increment $\alpha \approx 0.19$ mL g$^{-1}$, the assumed $\Delta n$ would have been $14\%$ smaller and the retrieved volume inflated to $\approx 103$ fL, outside the normal range. The erythrocyte therefore validates the geometry and the index convention at once, and illustrates that in phase volumetry the dominant uncertainty is usually not optical but the value adopted for $\Delta n$.

Neither the TIE nor the biological standard improves lateral resolution, which remains bounded by the effective aperture of the lens-free geometry. What they provide are error channels orthogonal to the reconstruction: an axial scale imposed mechanically rather than inferred, and a morphology whose reference values were established by entirely different techniques. A topography confirmed along both cannot be an artefact of the twin-image term, the dominant systematic of single-shot in-line holography \cite{latychevskaia2015,Zuo2020}.

\section{Conclusions}
This work demonstrates the effectiveness of DLHM for the 3D characterization of selective Chilean pollens, integrating the Kirchhoff-Helmholtz formalism with low-cost optoelectronic hardware to establish a baseline for automated, label-free melissopalynology.

Three elements turn that baseline into a measurement rather than a visualisation, each testing what the previous cannot. Certified polystyrene microspheres make the source--sample distance an optically retrieved quantity instead of a mechanical guess, with a volumetric accuracy of $1.3\%$ against an analytical ground truth; since standard and specimen share the same chamber and optical path, the calibration transfers to any field deployment. A human erythrocyte reconstructed with a haemoglobin refractive-index increment taken from the literature \cite{Gienger2019} reproduces the Evans--Fung discocyte profile and returns $88.9$ fL, none of it fitted, extending the validation to the low-contrast regime where biological specimens live. And the Transport-of-Intensity retrieval on bright-field micrographs, on a lateral scale traceable to a resolution standard \cite{Sanhueza2026}, reproduces the morphology of \textit{G. avellana} on an independent instrument and returns an area consistent with the value reported for thousands of expert-verified masks of that species. That channel is used for what its calibration supports, namely the lateral scale and the projected morphology, and not for volumetry: its defocus is a reproducible instrumental step, not a distance matched to the grain, so no absolute height profile follows from it.

Simulated and experimental reconstructions of light volumes around the focal plane remain a complementary route to estimating real-world distance and volume \cite{Staforelli2010,Lencina2012,bhabhor2024}. Future work will extend the standard-based protocol to multi-size sphere sets, exploit the KH--TIE discrepancy as a classification feature, and pursue real-time classification with the deep-learning frameworks now established \cite{kim2025,oconnor2020,Wu2019}. Applications follow in floral certification and adulteration detection in honeys \cite{Machuca2022,Jofre2025}, and in erythrocyte dynamics under physiological and hyperglycemic conditions, where segmented 3D maps from single holograms complement existing measurements \cite{Tapia2021,Bravo2023,Enriquez2026}.

\section*{Acknowledgments}
This work was supported by FONDEF Project IT24I0064 and the Facultad de Ciencias Físicas y Matemáticas, Universidad de Concepción (UDEC). The authors gratefully acknowledge the Laboratory of Palynology and Plant Ecology, School of Sciences and Technologies, University of Concepción, Los Angeles Campus, for the botanical identification of the collected material, the acetolysis protocol and the reference pollen collection on which this work relies.

\end{document}